\documentstyle[preprint,prb,aps,fancyhdr]{revtex}

\begin{document}

\draft
\rhead{let's try}
\lhead{let's try}
\title{Imaging Microwave Electric Fields Using a Near-Field Scanning
       Microwave Microscope}

\author{S. K. Dutta, C. P. Vlahacos, D. E. Steinhauer, Ashfaq S. Thanawalla,
        B. J. Feenstra, \mbox{F. C. Wellstood}, and
        Steven M. Anlage\footnote{Electronic mail: anlage@squid.umd.edu}}
\address{Center for Superconductivity Research, Department of Physics,
         University of Maryland, \mbox{College Park, Maryland 20742-4111}}

\author{Harvey S. Newman}
\address{Naval Research Laboratory, Washington, DC 20375}

\date{\today}
\maketitle

\begin{abstract}
By scanning a fine open-ended coaxial probe above an operating microwave
device, we image local electric fields generated by the device at microwave
frequencies.  The probe is sensitive to the electric flux normal to the face
of its center conductor, allowing different components of the field to be
imaged by orienting the probe appropriately.  Using a simple model of the
microscope, we are able to interpret the system's output and determine the 
magnitude of the electric field at the probe tip.  We show images of electric field 
components above a copper microstrip transmission line driven at 8 GHz, 
with a spatial resolution of approximately 200 $\mu$m.
\end{abstract}



\narrowtext

Simulating the electromagnetic behavior of passive microwave devices provides 
an attractive route toward developing new and improved devices. Ultimately, 
the information obtained from such simulations includes
the scattering parameters and the direction and
magnitude of the electric fields near the device. 
To check the reliability of these results, one can go beyond standard s-parameter
measurements, and use imaging techniques to determine 
the electromagnetic fields experimentally. 
Methods for imaging local electric fields include
using modulated scatterer probes,\cite{budka} coaxial cable probes,
\cite{ygao,cgao,vlahacos,anlage} electrooptic sampling,\cite{hou,weingarten}
extensions of atomic force microscopy,\cite{weide} and scanning SQUID 
microscopy.\cite{GuyEfield}  
Most of these techniques require expensive components and complicated configurations.
In this Letter, we describe the use of a relatively simple technique, employing an
open-ended coaxial probe to image the vector components of the local electric
field generated by operating microwave circuits. In addition, we discuss
how to interpret the images in terms of electric fields present at the face of the probe.

Our experimental configuration is shown schematically in Fig.\ \ref{parts}.%
\cite{vlahacos}  We use a coaxial cable probe with a center conductor diameter
$d_{c} = 200\ \mu\mbox{m}$ and an outer diameter $d_{o} = 860\ \mu\mbox{m}$.
During a scan, the probe is held at a constant height $h$ above the sample,
typically between 10 $\mu$m and several millimeters.  Rf electric fields from
the sample induce a high-frequency potential difference between the center and
grounded outer conductors of the probe.  The probe is connected to the input
port of a directional coupler, which has a length of coaxial cable attached to
its output port, thereby creating a resonant circuit.  The coupled port of the
directional coupler is connected to a matched diode detector, which produces a
voltage output proportional to the incident rf power.  The diode voltage is
low-pass filtered, amplified, and recorded by a computer.  The computer also
controls a two-axis translation stage, which raster scans the sample
underneath the probe.

The relationship between the diode voltage output and the electric field at
the probe face can be found by analyzing a circuit model for the system.  The
main idea is that the charge $Q$ induced on the exposed face of the probe's
center conductor is proportional to the integral of the normal component of
the electric field $E_{n}$ over this face.  We note that this naturally limits
the spatial resolution of the technique to no better than the diameter of the
center conductor.  The induced current at the probe face is then $I = \dot{Q}
= i \omega \epsilon_{o} E_{n} A$, where A is the area of the center conductor
face and $\omega$ is the angular frequency of the microwave field.

In order to relate $I$ to the voltage in the probe-coupler-cable assembly,
we need to know the impedance $Z_{m}$ that this microscope assembly presents
to an input signal.  Using standard transmission line theory,\cite{ramo} we
find:
\begin{equation}
   Z_{m} = Z_{o} \left[ \frac{(Z_{c} + Z_{o}) + (1-\delta)^{2} (Z_{c} - Z_{o})}
                             {(Z_{c} + Z_{o}) - (1-\delta)^{2} (Z_{c} - Z_{o})}
                 \right],
\label{Zm}
\end{equation}
where $Z_{o} = 50\ \Omega$ is the characteristic impedance of the transmission
line, $Z_{c}$ is the input impedance of the resonant cable,\cite{ramo} and
$\delta$ is the fraction of the input signal voltage that the directional
coupler taps off to its coupled port.  For the parameters of the resonant
section used (2 m long, open-ended cable with an attenuation of 1.1 dB/m,
driven at resonance near 8 GHz), $Z_{c} \approx 200\ \Omega$.  As a result,
for our -10 dB ($\delta \approx 0.3$) directional coupler, one finds
$Z_{m} \approx 90\ \Omega$.

By equating the current in the cable assembly to the current generated by the
field, we can solve for the rms magnitude of the field at the probe in terms
of the measured diode voltage:
\begin{equation}
   |E_{n}| = \frac{2}{\omega \epsilon_{o} A \delta |Z_{m} + Z_{o}|} D(V_{dc}),
\label{En}
\end{equation}
where $D$ is the high-frequency rms voltage at the diode input corresponding
to the measured diode output voltage $V_{dc}$.

The use of a coaxial probe geometry has several consequences for the correct
interpretation of Eq.\ (\ref{En}).  First, the expression is valid provided
we take $E_{n}$ to represent the field at the probe face when the probe is
present.  In general, this field may differ significantly from the field
$E_{n}^{o}$ in the absence of the probe.  The nature of the perturbation
due to the probe is discussed in more detail below.  Second, the derivation
of Eq.\ (\ref{En}) assumes that the coupling between the coaxial probe and
the device primarily occurs via the exposed face of the center conductor,
\textit{i.e.} only the field normal to this face is detected.  Because the
outer conductor shields the transverse components of the electric field,
this assumption is justified.  This shielding means we can image individual
components of the field by orienting the face of the probe in the appropriate
direction. The inset in the lower right of Fig.\ \ref{parts} shows two such
orientations, denoted ``vertical'' (probe face in the xy-plane) and
``horizontal'' (probe face in the yz-plane).

To investigate the capabilities of the microscope, we imaged a simple copper
microstrip transmission line (see Fig.\ \ref{parts}).  The microstrip consists
of a ground plane and a 2 mm wide and 45 mm long strip which are each 30
$\mu$m thick and separated by a 1.5 mm thick dielectric ($\epsilon_{r}
\approx 3.55$).  We connected a microwave source directly to one end of the
strip while the opposite end was left open, so that standing waves form with
a voltage antinode at the open end.

Figure \ref{topzx}(a) shows an image of a single antinode in the middle of
the strip, where the dashed lines indicate the edges of the strip.%
\cite{back}  Here, the probe was in the vertical orientation with
$h = 25\ \mu\mbox{m}$ and the source frequency was 8.05 GHz.  The wavelength
of the standing wave pattern was observed to decrease with increasing source
frequency, as expected.\cite{anlage}  We note that the field peaks strongly
just within the edges of the strip and that there are weak lobes away from
the strip.

Orienting the probe horizontally with the probe face pointing in the negative
\^{x} direction (as in the lower right inset of Fig.\ \ref{parts}) with
$h = 455\ \mu\mbox{m}$ results in a rather different image [see Fig.\
\ref{topzx}(b)].  In contrast to Fig.\ \ref{topzx}(a), note that the peak at
$\mbox{X}>0$ in Fig.\ \ref{topzx}(b) is located well outside the strip.
Furthermore, notice that the peak at $\mbox{X}<0$ is much weaker.  
Compared to the vertical configuration, in the horizontal orientation, the outer conductor 
more strongly perturbs the field when
situated over the strip, thereby decreasing the signal and causing an
asymmetry in the final image.  In order to accurately image the peak at
$\mbox{X}<0$, the probe must be rotated such that its face points in the
positive \^{x} direction, \textit{i.e.}\ the mirror image of the orientation
in the inset to Fig.\ \ref{parts}.  

The most important issue, however, is the remarkable difference in the signal
measured for different probe orientations. Figure \ref{vert/line}(a,b) shows that
this difference is also evident when the sample is scanned in the xz-plane.
For example, the field decreases more rapidly with increasing height for the 
vertical probe than the horizontal probe (note that for the horizontal probe 
orientation, the smallest probe/sample separation is limited
by the outer conductor radius).

The fact that the images depend strongly on probe orientation is consistent
with the probe being sensitive to individual components of electric field.
To compare the data with the expected field profile at a standing wave
antinode [\textit{i.e.}\ a slice in the xz-plane, such as in Fig.\
\ref{vert/line}(a,b)], we model the microstrip as a conductor held at a
constant potential above a grounded plane.  We then find the electric field
by numerically solving Poisson's equation in the inhomogeneous medium
surrounding the conductors and averaging the result over the area of the
probe's center conductor.

Naturally, we expect only qualitative agreement between the static model and
the data.  First of all, the model is not a full-wave solution at microwave
frequencies for inhomogeneous microstrip.\cite{hoffman}  In addition, it
neglects the presence of the microscope.  For instance, the grounded outer
conductor of the probe causes a local redistribution and screening of electric
field (more pronounced when using a horizontal probe).  Also, the current
delivered by the sample to the microscope will depend on its capacitive
coupling $C_{p}$ to the probe and on the value of $Z_{m}$.  In the unperturbed
case, the potential on a small segment of the strip sees a path to ground via
the vacuum.  However, with the probe present, the voltage is dropped across
the probe/sample impedance $Z_{p} = 1/i \omega C_{p}$ and $Z_{m}$ (see inset
in upper left of Fig.\ \ref{parts}).  Clearly, the value of these two
impedances will have a significant effect on the measured electric field.

Figure \ref{vert/line}(c) shows constant-height line scans above the strip for
the experimental data (solid) and model calculations (dotted) for the vertical
($E_{z}$) probe orientation.  Figure \ref{vert/line}(d) shows the
corresponding data for the horizontal ($E_{x}$) probe orientation.  In order
to make the magnitude of the fields comparable, the simulation values have
been scaled up by a factor of 65 for the vertical and 25 for the horizontal
orientation. We note that despite the simplifying assumptions made in our
model, the experimental results reproduce three key features in the
simulations which are consistent with the imaging of separate field
components: (i) $E_{z}$ is large at the center of the strip, while $E_{x}$ is
nearly zero; (ii) $E_{z}$ peaks just inside the strip edge, while $E_{x}$
peaks outside; and (iii) $|E_{z}|$ has a minimum just to the side of the
strip, while $E_{x}$ dies off slowly and monotonically to zero, away from the
strip.

Although the experimental and model curves are qualitatively similar, there
are large quantitative differences.  This is largely due to the fact that the
perturbation caused by the probe has not been taken into account in the
static electric field model.  To make a quantitative comparison between
experiment and theory, it is essential to calculate the field with the probe
present.  By knowing the values of $Z_{p}$, $Z_{m}$, and the voltage at some
point on the strip, we can find the voltage drop across $Z_{p}$ and directly
determine the value of the perturbed electric field at that location.
Treating the probe/sample capacitance $C_{p}$ as a parallel plate capacitor
yields $C_{p} \approx 6\ \mbox{fF}$ ($|Z_{p}| \approx 3.2\ \mbox{k} \Omega$
at 8 GHz) for a vertical probe at $h = 45\ \mu\mbox{m}$.\cite{compimp}  
From independent
measurements, the rms voltage at the center of an antinode [\textit{e.g.}\
X = 0, Y = 0 in Fig.\ \ref{topzx}(a)] was found to be 1.25 V.  With these
values, we estimate $|E_{z}| \approx 27\ \mbox{V/mm}$.  In comparison, the
experimental value is 62 V/mm and the simple model yields 0.6 V/mm.  Thus, the
measured electric field is enhanced in a manner consistent with our
expectations.

The data presented in Figs.\ \ref{topzx} and \ref{vert/line} have an
uncertainty of approximately 1\% for the peak values and 4\% for the regions
far from the strip.  However, uncertainty in probe height ($\sim$ 5 $\mu$m)
may give rise to a large effective decrease in the precision, especially if
the field is strongly dependent on height.  From Eq.\ (\ref{En}), one sees
that the diode voltage scales roughly as the area of the center conductor for
a given electric field.  Therefore at 8 GHz, for the $d_{c} = 200\
\mu\mbox{m}$ probe used here, the microscope works well for field strengths
greater than $0.5$ V/mm, while a larger probe ($d_{c} = 510\ \mu\mbox{m}$)
would operate precisely above 0.1 V/mm.  The electric field sensitivity is mainly limited by
the sensitivity of the electronics and the 20 ms averaging time. It could be 
significantly enhanced by using a low-noise amplifier or phase-sensitive 
detection of an amplitude-modulated signal.
However, the present sensitivity is already sufficient for the application of this imaging
technique to, for instance, superconducting filters.\cite{ashfaq}

In conclusion, using a near-field scanning microwave microscope, we have
demonstrated the imaging of individual components of electric field above a
microstrip line.  The microscope is composed of simple components and has a
wide range of operation (about 80 MHz to 50 GHz).  Although the coaxial probe
perturbs the fields it is measuring, the mechanism for this perturbation is
understood.  In principle, it should be possible to correct for this effect.
Finally, we note that scanning can be done fairly quickly; Fig.\ \ref{topzx}
was sampled at a rate of 50 Hz with a spatial resolution of 50 $\mu$m,
yielding an image in about 15 minutes.

The authors wish to acknowledge useful discussions with D. W. van der
Weide.  This work has been supported by NSF grant \# ECS-9632811, NSF-MRSEC
grant \# DMR-9632521, and by the Maryland Center for Superconductivity
Research.



\begin{figure}
\caption{Schematic showing the main components of the microwave microscope.
   An open-ended coaxial probe is brought near an operating device, so that
   stray electric fields induce a microwave signal in the probe.  The inset
   in the lower right shows probe orientations used to measure different field
   components, at a common probe/sample separation $h$.  The inset in the
   upper left shows the interaction of the device with the microscope.}
\label{parts}
\end{figure}

\begin{figure}
\caption{Images of two components of electric field taken above a 2 mm wide
   copper microstrip driven at 8.05 GHz.  The dashed lines show the boundaries
   of the strip. (a) Image with vertical probe ($h = 25\ \mu\mbox{m}$);
   contour lines at 70, 50, 30, 10, 3 V/mm.  (b) Image with horizontal probe
   ($h = 455\ \mu\mbox{m}$); contour lines at 11, 7.5, 4, 1.5 V/mm.}
\label{topzx}
\end{figure}

\begin{figure}
\caption{Electric field in the xz-plane above the microstrip (at Y = 0 in
   Fig.\ \ref{topzx}) with the probe (a) vertical (contour lines at 80, 55,
   30, 15, 8 V/mm) and (b) horizontal (contour lines at 12, 9, 6, 4 V/mm).
   The strip is shown as a hatched rectangle on the x-axes.  Solid lines in
   (c) and (d) show line cuts (at Y = 0) of experimental data at two probe
   heights, as labeled in $\mu$m.  The dotted lines show the corresponding
   numerical field simulation, neglecting probe perturbation.  The vertical
   simulation data are multiplied by 65 and the horizontal by 25.  The edge of
   the strip is indicated by the dashed vertical lines.}
\label{vert/line}
\end{figure}


\begin{references}
\bibitem{back}A small constant offset was subtracted from all plots, on
  the basis that the field should vanish far from the strip.
\bibitem{compimp} For Z$_p = 3.2\;{\mbox k}\Omega$ and Z$_m = 90\;\Omega$, 
even substantial changes in Z$_c$ and $\delta$ will not significantly affect 
the value of the measured electric field.
\end{references}
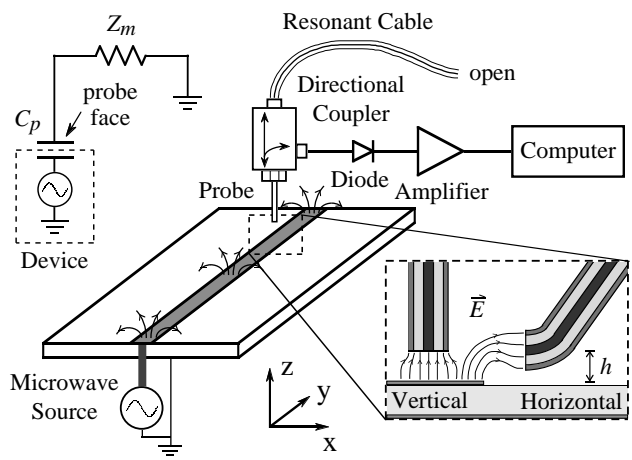
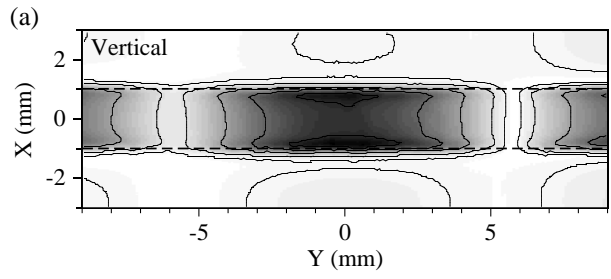
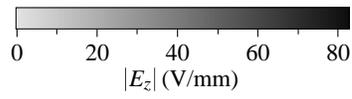
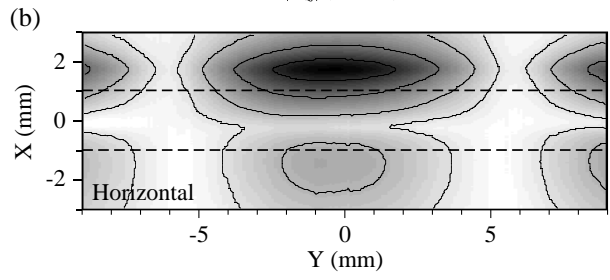
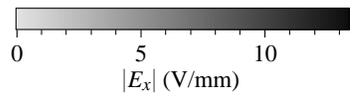
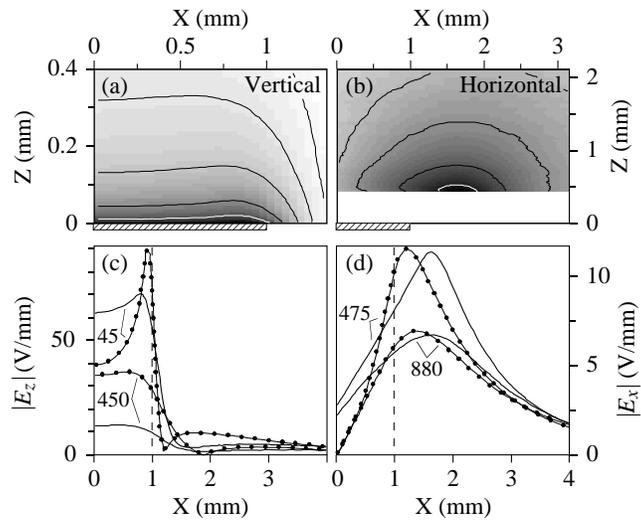
\end{document}